\documentclass{article}
 \renewcommand{\a}{\alpha}
 \renewcommand{\b}{\beta}

 \newcommand{\ad}{a^{\dag}}
 \newcommand{\p}{\prime}
 \newcommand{\ap}{\alpha^{\p}}
 \renewcommand{\sp}{s^{\p}}
 \newcommand{\bp}{\beta^{\p}}
 \newcommand{\tp}{t^{\p}}
 
 \newcommand{\bd}{b^{\dag}}

 \newcommand{\asbt}{\alpha s,\beta t}
 \newcommand{\asbtp}{\alpha^{\p} s^{\p},\beta^{\p} t^{\p}}
 \usepackage{graphics}
          \begin{document}
           \title{Complex Rational Numbers in Quantum mechanics}
          \author{Paul Benioff\\
           Physics Division, Argonne National Laboratory \\
           Argonne, IL 60439 \\
           e-mail: pbenioff@anl.gov}
           \date{\today}
           \maketitle
           \begin{abstract}
           A binary representation of complex rational
           numbers and their arithmetic is described that is
           not based on qubits. It
           takes account of the fact that $0s$ in a qubit string
           do not contribute to the value of a number. They
           serve only as place holders. The representation is
           based on the distribution of four types of systems,
           corresponding to $+1,-1,+i,-i,$ along an integer
           lattice. Complex rational numbers correspond to arbitrary
           products of four types of creation operators acting on the vacuum
           state. An occupation number representation is given for
           both bosons and fermions.
           \end{abstract}

           \section{Introduction}
           Quantum computation and quantum information continue to
           attract much interest and study.  Much of the interest
           was stimulated by work showing that quantum computers
           could solve some problems  more efficiently than any
           known classical computer \cite{Shor,Grover}.  Also some
           recent work addresses the possible relation between
           quantum computing and questions in cosmology and
           quantum gravity \cite{Lloyd,Zizzi}.

            In all of this work qubits (or qudits for d-dimesional
          systems) play a basic role. As quantum binary systems
          the states $|0\rangle,|1\rangle$ of a qubit represent the binary
          choices in quantum information theory.  They also
          represent the numbers $0$ and $1$ as numerical inputs
          to quantum computers. For $n$ qubits, corresponding
          product states, such as $|\underline{s}\rangle=
          \otimes_{j=1}^{n}|s(j)\rangle$ where $s(j)=0 \mbox { or
          }1,$ represent a specific $n$ qubit information state.
          These states and their linear superpositions are
          inputs to quantum computers.

          Even though representation of numbers as strings of
          qubits or as strings of bits in classical work is
          widespread \cite{Nielsen}, it is not essential. Other representations
          are possible and may be useful.  One of these is based
          on the observation that the $0s$ in a qubit or bit
          string serve only as place holders.  They do not
          contribute to the value of the number.

          This suggests a representation that does not use qubits or bits.
          It is based on the distributions of $1s$ along an integer
          lattice.  For example the rational binary number
          $10100.0011$ would be represented here as
          $1_{4}1_{2}1_{-3}1_{-4}.$

         This will be taken over to quantum mechanics by
         representing complex rational numbers as states of
         systems on a discrete lattice. The representation will be
         based on the use of annihilation and
         creation operators that create and annihilate systems on
         a lattice.  Such an approach is useful in cases where
         particle numbers are not conserved. This is the case here
         as arithmetic operations do not conserve the number of $1s$ in
         numbers.\footnote{They also do not conserve bit string lengths.
         This is usually accounted for by use of truncation to a
         fixed accuracy dependent length in computations.}

         Both bosons and fermions will be considered.  For
         bosons the basic creation operators are
         $\ad_{\a,j}$ and $\bd_{\b,k}.$ The state $|0\rangle$ is the vacuum
         state. The single particle states $\ad_{\a,j}|0\rangle$ and
         $\bd_{\b,k}|0\rangle$ show a type $a$ system in internal state $\a$
         at lattice site $j$ and a type $b$ system in internal state $\b$
         at site $k$. They correspond respectively to the
         real and imaginary numbers $\a2^{j}$ and $i\b2^{k}$ where
         $\a=+,-$ and $\b=+,-$ denote the sign of the numbers.
         Multiple particle states and linear superpositions of these states
         are described using products of these operators, as in
         $1/\sqrt{2}(\ad_{+,7}\ad_{-,6}\bd_{-,4}|0\rangle+
         \ad_{-,-2}\bd_{-,6}|0\rangle).$

         For fermions an additional discrete index $h=1,2\cdots$
         is needed. This index does not contribute to the
         numerical value of a state but it does allow such states
         as $\ad_{\a,j,1}\ad_{\a,j,2}|0\rangle$ which corresponds to
         the number $\a(2^{j}+2^{j}).$ Such states arise naturally
         during arithmetic operations.

         This work differs from other work on fermionic quantum
         computation \cite{Kitaev,Ozhigov} and number
         representation in quantum mechanics \cite{BenRNQMALG} in
         that it is not based on logical or physical qubits. It
         also differs in the use of an occupation number
         representation which results in both standard and
         nonstandard representations of complex rational numbers.
         These representations are defined in the next section.

         The use of different $a$ and $b$ systems to represent
         real and imaginary numbers is done here to help
         understanding.  It is completely arbitrary in that one
         can also use just one type of system with an extra
         internal state index as in $\ad_{\a,x,j}$ (bosons) or
         $\ad_{\a,x,j,h}$ (fermions) where $x=r,i.$

         Here complex rational numbers are represented by
         all finite products of creation operators acting on the
         vacuum state $|0\rangle$.  These states and their linear superpositions
         form a Fock space $\mathcal H^{Ra}$ of states.  This
         representation is quite compact in that all four types of
         numbers can be included in one state.  For example, the
         boson state $\ad_{+,2}\ad_{-,0}\bd_{-,3}\bd_{+,-1}\ad_{-,-2}|0\rangle,$
         which represents the number $2^{2}-2^{0}-2^{-2}+i(-2^{3}+2^{-1}),$ has
         the qubit representation $|10.11\rangle,|-i111.1\rangle.$

         This flexibility and
         compactness allows a representation of many positive and
         negative  complex rational numbers, which are to be
         combined into one complex number, as a single operator
         product acting on the vacuum.  Such collections or
         matrices of numbers can occur, for example, in evaluating
         integrals of complex functions. Here one may collect many
         values of a function $f(x)$ over a range $u<v$ which are
         combined to evaluate the integral $\int_{u}^{v}f(x)dx.$

         In the next section basic properties of the annihilation creation (a-c) operators
         and their use in rational number states are
         outlined for both bosons and fermions. The most general
         states, which can have more than one system at a  $j$
         lattice site, and their reductions to standard complex
         rational number states are discussed. Section \ref{AO}
         summarizes basic arithmetic operations on the states and
         their relation to complex number equivalents in $C$ of
         rational numbers.  Here $C$ is the complex number field
         over which $\mathcal H^{Ra}$ is defined. More details
         on complex rational number states are given in
         \cite{BenRCRNQM}.

         It should be emphasized that rational number states and
         their arithmetic operations (addition, subtraction,
         multiplication and division to arbitrary accuracy)
         will be described here with
         no reference to the numbers in $C$ and their arithmetic
         properties.  An operator $\tilde{N}$ will be described
         that associates a complex rational number in $C$ to each
         complex rational number state. The fact that $\tilde{N}$
         is a morphism and preserves arithmetic properties is
         satisfying, but it plays no role in defining the
         properties of and operations on the states.

         Finally one should note that the states described here as
         complex rational states do not correspond to all
         rational number states. However they are dense in the
         rational number states  and can approximate any rational
         number state to arbitrary accuracy.  For example, no rational
         state defined here corresponds to the number $1/3.$  However
         the rational states correspond to numbers that approximate
         $1/3$ to arbitrary accuracy. They also correspond to the types
         of numbers used by computers in actual calculations.

         \section{Complex Rational Number States} \label{CRNS}

         One begins with the commutation relations for the basic
         a-c operators.  For bosons one has \begin{equation}\label{boscomm}
         \begin{array}{c}[a_{\a ,j},a_{\ap,k}]=
         [b_{\b ,j},b_{\bp,k}]=[\ad_{\a,j},\ad_{\ap,k}]=
         [\bd_{\b,j},\bd_{\bp,k}]=0 \\ \mbox{$[a_{\a,j},\ad_{\ap,k}] =
         \delta_{\a,\ap}\delta_{j,k}$};\;\;\;\;\mbox{$[b_{\b,j},
         \bd_{\bp,k}] = \delta_{\b,\bp}\delta_{j,k}$}.\end{array}
         \end{equation} For fermions the anticommutation relations
         are \begin{equation}\label{fercomm}
         \begin{array}{c}\{a_{\a ,j,h},a_{\ap,k,h^{\p}}\}=
         \{b_{\b ,j,h},b_{\bp,k,h^{\p}}\}=\{\ad_{\a,j,h},\ad_{\ap,k,h^{\p}}\}=
         \{\bd_{\b,j,h},\bd_{\bp,k,h^{\p}}\}=0 \\ \mbox{$\{ a_{\a,j,h},\ad_{\ap,k,h^{\p}}\} =
         \delta_{\a,\ap}\delta_{j,k}\delta_{h,h^{\p}}$};\;\;\;\;\mbox{$\{ b_{\b,j,h},
         \bd_{\bp,k,h^{\p}}\} = \delta_{\b,\bp}\delta_{j,k}\delta_{h,h^{\p}}$}\end{array}
         \end{equation} where $\{c,d\}=cd+dc.$  The $a$ and $b$
         operators commute for both bosons and fermions as
         they represent distinguishable systems.

          A complete basis set of states can be defined in terms of
         occupation numbers of the various boson or fermion states.
         Let $n_{+},n_{-},m_{+},m_{-}$
         be any four functions that map the set of all integers to the
         nonnegative integers. Each function has the value $0$
         except possibly on a finite set of integers. Let
          $s,s^{\p},t,t^{\p}$ be the four finite sets of integers
          which are the nonzero domains, respectively, of the four
          functions. Thus $n_{+,j}\neq 0[=0]$ if $j\epsilon s[j
          \mbox{ not in }s],$  $n_{-,j}\neq 0[=0]$ if $j\epsilon s^{\p}
          [j \mbox{ not in }s^{\p}],$ etc.

         Let $\bigcup s,t$ be the set of all integers in one or more of the four sets.
         Then a general boson occupation number state has the form
         \begin{equation}\label{occno}
         |n_{+},n_{-},m_{+},m_{-}\rangle = \prod_{j\epsilon \cup
         s,t}|n_{+,j},n_{-,j}m_{+,j}m_{-,j}\rangle\end{equation}
         where $|n_{+,j},n_{-,j}m_{+,j}m_{-,j}\rangle,$ the
         occupation number state for site $j,$ is given by
         \begin{equation}\label{occnost}\begin{array}{l}
         |n_{+,j},n_{-,j}m_{+,j}m_{-,j}\rangle=\frac{1}{N(n,m,+,-,j)} \\
         \hspace{1cm}\times (\ad_{+,j})^{n_{+,j}}
         (\ad_{-,j})^{n_{-,j}}(\bd_{+,j})^{m_{+,j}}(\bd_{-,j})^{m_{-,j}}|0\rangle.
         \end{array}\end{equation}  The normalization factor
         $N(n,m,+,-,j)=(n_{+,j}!n_{-,j}!m_{+,j}!m_{-,j}!)^{1/2}.$
         Note that the product $\prod_{j\epsilon \cup
         s,t}$ denotes a product of creation operators, and not
         a product of states.

         The equivalent fermionic representation for the
         state $|n_{+},n_{-},m_{+},m_{-}\rangle$ is based
         on a fixed ordering of the a-c operators.
         In this case the product $(\ad_{+,j})^{n_{+,j}}$ becomes
         $\ad_{+,1,j}\cdots\ad_{+,h,j}\cdots\ad_{+,n_{+,j},j}$
         with similar replacements for $(\ad_{-,j})^{n_{-,j}},$
         $(\bd_{+,j})^{m_{+,j}},$ and $(\bd_{-,j})^{m_{-,j}}.$
         Each component state
         $|n_{+,j},n_{-,j},m_{+,j},m_{-,j}\rangle$
         in Eq. \ref{occnost} is given by
         \begin{equation}\label{occferm}\begin{array}{c}
         |n_{+,j},n_{-,j},m_{+,j},m_{-,j}\rangle=
         \ad_{+,n_{+,j},j}\cdots \ad_{+,1,j}\ad_{-,n_{-,j},j}\cdots \\
         \ad_{-,1,j}  \bd_{+,m_{+,j},j}\cdots\bd_{+,1,j}
         \bd_{-,m_{-,j},j}\cdots\bd_{-,1,j}|0\rangle\end{array}\end{equation}
         The final state is given by an ordered product over the $j$ values,
         \begin{equation}\label{occnoferm}|n_{+},n_{-},m_{+},m_{-}\rangle =
         \prod_{j\epsilon \cup s,t}J|n_{+,j},n_{-,j}m_{+,j}m_{-,j}
         \rangle.\end{equation} Here $J$ denotes a $j$ ordered
         product where factors with larger values of $j$ are to
         the right of factors with smaller $j$ values. The choice
         of ordering, such as that used here in which
         the ordering of the $j$ values is the opposite of that
         for the $h$ values which increase to the left as in Eq.
         \ref{occferm}, is arbitrary.  However, it must remain fixed
         throughout.

            The interpretation of these states is that they are the
         boson or fermion equivalent of \emph{nonstandard} representations of
         complex rational numbers as distinct from \emph{standard}
         representations.\footnote{This use of standard and nonstandard is
         completely different from standard and nonstandard numbers
         described in mathematical logic \cite{Chang}.} Such nonstandard
         states occur often in arithmetic operations.
         They correspond to columns of binary numbers where
         each number in the
         column is any one of the four types, positive real,
         negative real, positive imaginary, and negative
         imaginary. In a boson representation, individual systems
         are not distinguishable. The only measurable properties
         are the number of systems of each type  $+1,-1,+i,-i$ in
         the single digit column at  each site $j.$ Individual
         systems are distinguishable in a fermion representation.
         However the variable $h$ that separates fermions with the
         same value of $\a$ and $j$ does not contribute to the
         numerical value of the state.

         An already mentioned example of a nonstandard representation
         is that made by a computation of the value
         of the integral $\int_{u}^{v}f(x)dx$
         of a complex valued function  $f.$ The table,
         or matrix, of $M$ results obtained by computing in parallel,
         or by a quantum computation, values of $f(x_{\ell})$ for
         $\ell=1,2,\cdots,M$ is represented here by a state
         $|n_{+},n_{-},m_{+},m_{-}\rangle$ where
         $n_{+,j},n_{-,j}m_{+,j},m_{-,j}$ give the number of
         $+1's$, $-1's,$ $+i's$, and $-i's$ in the column at site $j.$ This is
         a nonstandard representation because it is numerically
         equal to the final result which is a standard
         representation consisting of one real and one imaginary rational
         number, often represented as a pair, $w,iy$.

         Most of the occupation number states are nonstandard
         representations.  The standard representations are
         characterized by the restrictions that at most one of
         $n_{+},n_{-}$ and one of $m_{+},m_{-}$ have nonempty
         domains and that the functions have the constant value
         $1$ on their domains.  The four possibilities are \begin
         {equation}\label{stdbos}\begin{array}{c}
         |\underline{1}_{s},0,\underline{1}_{t},0\rangle =(\ad_{+})^{s}
         (\bd_{+})^{t}|0\rangle \\|\underline{1}_{s},0,0,
         \underline{1}_{t^{\p}}\rangle =(\ad_{+})^{s}
         (\bd_{-})^{t^{\p}}|0\rangle\\|0,\underline{1}_{s^{\p}},
         \underline{1}_{t},0\rangle = (\ad_{-})^{s^{\p}}(\bd_{+})^{t}|0\rangle
         \\|0,\underline{1}_{s^{\p}},0,\underline{1}_{t^{\p}}\rangle =
         (\ad_{-})^{s^{\p}}(\bd_{-})^{t^{\p}}|0\rangle.
         \end{array}\end{equation} Here $(\ad_{+})^{s}=\prod_{j\epsilon
         s}\ad_{+,j}$ and $\underline{1}_{s}$ denotes the constant
         $1$ function on $s$, etc. Pure real or imaginary  standard
         rational states are included if $t,\;t^{\p}$ or
         $s,\;s^{\p}$ are empty.  If $s,\;s^{\p},\; t,\; t^{\p}$ are
         all empty one has the vacuum state $|0\rangle.$  Note that
         Eq. \ref{stdbos} is also  valid for fermions  with the
         replacements\begin{equation}\label{stdfer}\begin{array}{l}
        (\ad_{\a} )^{s}\rightarrow \ad_{\a,1,j_{1}}\ad_{\a,1,j_{2}}
        \cdots\ad_{\a,1,j_{|s|}} \\ (\bd_{\b})^{t}\rightarrow
        \bd_{\b,1,k_{1}}\bd_{\b,1,k_{2}}\cdots\bd_{\b,1,k_{|t|}}.\end{array}
        \end{equation} Here $\a=+,-$, $\b=+,-$, and
        $s=\{j_{1},j_{2},\cdots,j_{|s|}\},\;
        t=\{k_{1},k_{2},\cdots,k_{|t|}\}$. Also
        $j_{1}<j_{2}<\cdots<j_{|s|},\;k_{1}<k_{2}<\cdots<k_{|t|},$
        and $|s|,|t|$ denote the number of integers in $s,t.$

        Standard states are quite important.  All theoretical
         predictions as computational outputs, and numerical
         experimental results are represented by standard real
         rational states. Nonstandard representations occur
         during the computation process and in any situation where
         a large amount of numbers is to be combined. Also qubit states
         correspond to standard representations only.

         This shows that it is important to describe the numerical
         relations between nonstandard representations and standard
         representations and to define numerical equality between states.
         To this end let \begin{equation}\label{Nequ}
         |n_{+},n_{-},m_{+},m_{-}\rangle
         =_{N}|n^{\p}_{+},n^{\p}_{-},m^{\p}_{+},m^{\p}_{-}\rangle\end{equation}
         be the statement that the two indicated states are numerically equal.
         Note that numerical equality has nothing to do with state equality
         in quantum mechanics. Two numerically equal states can be quite
         different physically.

         Numerical equality is defined by some basic requirements
         on a-c operators. For bosons they are
         \begin{equation}\label{abdjabj} \ad_{+,j}\ad_{-,j}=_{N}\tilde{1};\;\;\;\;
         \bd_{+,j}\bd_{-,j}=_{N}\tilde{1}\end{equation}
         and \begin{equation}\label{abjabj}\begin{array}{l} \ad_{\a,j}\ad_{\a,j}=_{N}
         \ad_{\a,j+1}\;\;\;\; a_{\a,j}a_{\a,j}=_{N}a_{\a,j+1} \\ \bd_{\b,j}\bd_{\b,j}=_{N}
         \bd_{\b,j+1}\;\;\;\;
         b_{\b,j}b_{\b,j}=_{N}b_{\b,j+1}.\end{array}\end{equation}
         For fermions one has
         \begin{equation}\label{fabdjabj}
         \ad_{+,j,h}\ad_{-,j,h^{\p}}=_{N}\tilde{1};\;\;\;\;
         \bd_{+,j,h}\bd_{-,j,h^{\p}}=_{N}\tilde{1}\end{equation}
         and \begin{equation}\label{fabjabj}\begin{array}{l}
         \ad_{\a,h,j}\ad_{\a,h^{\p},j}=_{N}
         \ad_{\a,h^{\p\p},j+1}\;\;\;\; a_{\a,h,j}a_{\a,h^{\p},j}
         =_{N}a_{\a,h,^{\p\p},j+1} \\ \bd_{\b,h,j}\bd_{\b,h^{\p},j}=_{N}
         \bd_{\b,h^{\p\p},j+1}\;\;\;\; b_{\b,h,j}b_{\b,h^{\p},j}
         =_{N}b_{\b,h^{\p\p},j+1}.\end{array}\end{equation}   In Eq. \ref{fabjabj}
         $h\neq h^{\p}.$ Otherwise the values of $h,h^{\p},h^{\p\p}\geq
         1$ are arbitrary except that removal of fermions is
         restricted to occupied $h$ values and addition is
         restricted to unoccupied values. To avoid poking holes in
         the successive values of $h$ at each site $j,$
         it is useful to restrict system removal to the maximum
         occupied h value and addition to its nearest unoccupied neighbor.
         However the $h$ values at which systems are added or
         removed do not affect the numerical value of the state.

         The first pair of equations says that any  state that
         has one or more $+$ and $-$ systems of either the $r$
         (real) or $i$ (imaginary) type at a site $j$ is
         numerically equivalent to the state with one less $+$ and
         $-$ system  at the site $j$ of either type. This is the
         expression here of $2^{j}-2^{j}=i2^{j}-i2^{j}=0$ for the
         numbers in $C$.
         The second set of two pairs, Eq. \ref{abjabj}, says that any
         state with two systems of the same type and in the same
         internal state at site $j,$ and two different $h$ values for
         fermions, is numerically equivalent to a state without these
         systems but with one system of the same type and internal state
         at site $j+1.$ This corresponds to $2^{j}+2^{j}=2^{j+1}$ or
         $i2^{j}+i2^{j}=i2^{j+1}.$

         From these relations one sees that any process whose
         iteration preserves $N$ equality according to Eqs.
         \ref{abdjabj} and \ref{abjabj} can be used to
         determine if Eq. \ref{Nequ} is valid for two different
         states. For example, for bosons if \begin{equation}\label{jj1}
         |n_{+},n_{-},m_{+},m_{-}\rangle
         =a_{+,j}a_{-,j}|n^{\p}_{+},n^{\p}_{-},
         m^{\p}_{+},m^{\p}_{-}\rangle\end{equation} or
         \begin{equation}\label{jjj+1}
         |n_{+},n_{-},m_{+},m_{-}\rangle
         =\bd_{+,j+1}b_{+,j}b_{+,j}|n^{\p}_{+},n^{\p}_{-},
         m^{\p}_{+},m^{\p}_{-}\rangle,\end{equation} then Eq.
         \ref{Nequ} is satisfied.

         One can use the a-c operators to define operators that
         carry out the changes on states implied by Eqs.
         \ref{abdjabj}, \ref{abjabj}, \ref{fabdjabj}, and
         \ref{fabjabj}. Explicit expressions are given in
         reference\cite{BenRCRNQM}.

         Reduction of a nonstandard representation to a standard
         one proceeds by iteration of steps based on the above
         equivalences.  At some point the process stops when
         the resulting state  has at most one system of the $a$ or
         $b$ type at each site $j$. This is the case for both
         bosons and fermions.  The possible options for each $j$
         can be expressed as \begin{equation}\label{stdconv}
         \begin{array}{l}|n_{+,j},n_{-,j},m_{+,j},m_{-,j}\rangle
         =\left\{ \begin{array}{l}|1,0,0,1\rangle \\
         |1,0,1,0\rangle\\|0,1,0,1\rangle\\|0,1,1,0\rangle\end{array}
         \mbox { or }\left\{\begin{array}{l}|0,0,0,1\rangle\\
         |0,0,1,0\rangle\\|1,0,0,0\rangle\\|0,1,0,0\rangle\end{array}\right.\right.
        \\ \mbox{} \\ \hspace{3cm} \mbox{ or }|0,0,0,0\rangle.
        \end{array}\end{equation} An example of such a
         state for several $j$ is $|1_{+,3}i_{+,3}1_{-,2}i_{-,4}
         1_{-,-6}\rangle.$ This state corresponds to the $C$ number
         $2^{3}-2^{2}-2^{-6}+i(2^{3}-2^{4}).$

         Conversion of a state in this form into a standard state requires
         first determining the signs of the $a$ and $b$ systems occupying the
         sites with the largest $j$ values. This determines the
         signs separately for the real and imaginary components of
         the standard representation.
         In the example given above the real component is $+$ as
         $3>2,-6$ and the imaginary component is $-$ as $4>3.$

         Conversion of all a-c operators into the same kind, as
         shown in Eq. \ref{stdbos}, is based on four relations
         obtained by iteration of Eq. \ref{abjabj} and use of
         Eq. \ref{abdjabj}.  For $k<j$  and for bosons
         they are \begin{equation}\label{abjk}
         \begin{array}{l}a^{\dag}_{+,j}\ad_{-,k} =_{N}a^{\dag}_{+,j-1}\cdots
         a^{\dag}_{+,k} \\ a^{\dag}_{-,j}\ad_{+,k}
         =_{N}a^{\dag}_{-,j-1}\cdots
         a^{\dag}_{-,k} \\ b^{\dag}_{+,j}\bd_{-,k}
         =_{N}b^{\dag}_{+,j-1}\cdots
         b^{\dag}_{+,k} \\ b^{\dag}_{-,j}\bd_{+,k}
         =_{N}b^{\dag}_{-,j-1}\cdots
         b^{\dag}_{-,k}.\end{array}\end{equation}  These equations
         are used to convert all $a$ and all $b$ operators to the
         same type ($+$ or $-$) as the one at the largest occupied
         $j$ value.  Applied to the example $|1_{+,3}i_{+,3}1_{-,2}
         i_{-,4}1_{-,-6}\rangle,$  gives
         $|1_{+,2}1_{+,1}1_{+,0}1_{+,-1}1_{+,-3}\cdots
         1_{+,-6}i_{-,3}\rangle.$ for the standard representation.

         The same four equations hold for fermions provided $h$
         subscripts are included. The values of $h$ are arbitrary
         as they do not affect $=_{N}.$ However, physically,
         application to a state of the form of Eq.\ \ref{stdconv}
         requires that $h=1$ everywhere, as in $a^{\dag}_{+,1,j}\ad_{-,1,k}
         =_{N}a^{\dag}_{+,1,j-1}\cdots a^{\dag}_{+,1,k}$ for
         example.

         \section{A Number Operator}\label{NO}

          It is  useful to define an operator $\tilde{N}$ that assigns to
         each complex rational state a corresponding
         complex rational number in $C$. Each standard and
         nonstandard complex rational state is an eigenstate
         of $\tilde{N}$. The eigenvalue for this state is the
         complex number in $C$ that $\tilde{N}$ associates with
         the state.

         For fermions  $\tilde{N}$ is defined by
         \begin{equation}\label{defN}\begin{array}{l}
         \tilde{N}=\sum_{h,j}2^{j}[\ad_{+,h,j}a_{+,h,j}-
         \ad_{-,h,j}a_{-,h,j} \\ \hspace{1cm}+i(\bd_{+,h,j}
         b_{+,h,j}-\bd_{-,h,j}b_{-,h,j})].\end{array}\end{equation}
         From this definition one can obtain the
         following properties:\begin{equation}\label{Ndef}\begin{array}{c}
          [\tilde{N},\ad_{\a,h,j}]
         =\a 2^{j}\ad_{\a,h,j} \\ \mbox{}[\tilde{N},\bd_{\b,h,j}]
         =i\b 2^{j}\bd_{\b,h,j} \\ \tilde{N}|0\rangle
         =0.\end{array}\end{equation} Here $\a = +,-$ and $\b =+,-.$
         These equations apply to bosons if the $h$ variable is
         deleted.

         The eigenvalues of $\tilde{N}$ acting on  states that are
         products of $\ad$ and $\bd$ operators can be obtained from
         Eqs.\ \ref{defN} or \ref{Ndef}. As an
         example, for the boson state $\ad_{+,k_{1}}\ad_{-,k_{2}}\bd_{+,k_{3}}\bd_{-,k_{4}}
         |0\rangle,$
         \begin{equation}\label{Nex}\begin{array}{l}
         \tilde{N}\ad_{+,k_{1}}\ad_{-,k_{2}}\bd_{+,k_{3}}\bd_{-,k_{4}}|0\rangle
         = \\ \hspace{0.5cm}(2^{k_{1}}-2^{k_{2}}+i2^{k_{3}}-i2^{k_{4}}) \\
         \hspace{1cm}\times\ad_{+,k_{1}}\ad_{-,k_{2}}\bd_{+,k_{3}}
         \bd_{-,k_{4}}|0\rangle.\end{array}\end{equation}
         For standard representations in general
         \begin{equation}\label{NNcs}\tilde{N}(\ad_{\alpha})^{s}(\bd_{\beta})^{t}
         |0\rangle =N[(\ad_{\alpha})^{s}(\bd_{\beta})^{t}](\ad_{\alpha})^{s}
         (\bd_{\beta})^{t}|0\rangle\end{equation} where \begin{equation}\label{Ncs}
         \tilde{N}[(\ad_{\alpha})^{s}(\bd_{\beta})^{t}]=\left\{\begin{array}
         {ll}2^{s}+i2^{t} & \mbox{ if } \alpha=+,\beta=+ \\ -2^{s}+i2^{t} &
         \mbox{ if } \alpha=-,\beta=+ \\ 2^{s}-i2^{t} & \mbox{ if } \alpha=+,\beta=-
         \\ -2^{s}-i2^{t} & \mbox{ if } \alpha=-,\beta=-.
         \end{array}\right.\end{equation} Here $2^{s}=\sum_{j\epsilon
         s}2^{j}$ and $2^{t}=\sum_{k\epsilon t}2^{j}$.

         These results also hold for fermion states. For standard
         states Eq. \ref{stdfer} gives an explicit representation
         for $(\ad_{\alpha})^{s}(\bd_{\beta})^{t}(\ad_{\alpha})^{s}
         (\bd_{\beta})^{t}|0\rangle.$

         The operator $\tilde{N}$ has the satisfying property that any
         two states that  are $N$ equal have the same  $\tilde{N}$
         eigenvalue. If the state $|n_{+},n_{-},m_{+},m_{-}\rangle=_{N}
         (\ad_{\alpha})^{s}(\bd_{\beta})^{t} |0\rangle$ then \begin{equation}
         \tilde{N}|n_{+},n_{-},m_{+},m_{-}\rangle =_{N}
         \tilde{N} (\ad_{\alpha})^{s}(\bd_{\beta})^{t}|0\rangle.\end{equation}
         This follows from Eqs. \ref{abdjabj},\ref{abjabj}, and \ref{defN}.

         These results show that the eigenspaces of $\tilde{N}$
         are invariant for any process of reducing a
         nonstandard state to a standard state using Eqs.
         \ref{abdjabj}-\ref{fabjabj}. Any state
         $|n_{+},n_{-},m_{+},m_{-}\rangle$ with $n_{+,j}\geq 1$ and
         $n_{-,j}\geq 1$ for some $j$ has the same $\tilde{N}$ eigenvalue as the
         state with both $n_{+,j}$ and $n_{-,j}$ replaced by
         $n_{+,j}-1$ and $n_{-,j}-1.$ Also if $n_{+,j}\geq 2$ then
         replacing $n_{+,j}$ by $n_{+,j}-2$ and $n_{+,j+1}$ by
         $n_{+,j+1}+1$ does not change the $\tilde{N}$ eigenvalue.
         Similar relations hold for $m_{-,j},m_{+,j},m_{-,j}.$
         These results show that each eigenspace of $\tilde{N}$
         is infinite dimensional. It is spanned by an infinite
         number of nonstandard complex rational  states
         and exactly one standard state.

         The usefulness of $\tilde{N}$ results from the fact that
         it is a morphism from the complex rational number basis
         in $\mathcal H^{Ra}$ to the complex rational numbers in
         $C$.  That is, it preserves arithmetic relations and
         operations.  It was also used implicitly in the preceding
         to supply numerical values to states as illustrations.

         It is important to note that $\tilde{N}$ is not
         used in any way to define the standard and nonstandard
         complex rational states or the basic properties of
         $=_{N}.$ It will also not be used in the next section to
         define and give properties of basic arithmetic
         operations. The definitions and arithmetic properties of
         the complex rational states stand on their own with no
         reference to $\tilde{N}$.  However, the operator can be used as
         a check to show that the arithmetic properties of the states
         are preserved by their $\tilde{N}$ images in $C$.

         \section{Arithmetic Operations}\label{AO}

         Here the definition and properties of arithmetic
         operations are limited to addition and subtraction.
         Also the discussion is limited to standard states and
         their linear superpositions. Extension to nonstandard states is
         straightforward as any state $N$ equal to a standard
         state has the same arithmetic properties as the standard
         state. Details on  nonstandard states and multiplication and
         division to any finite accuracy are given in reference\cite{BenRCRNQM}.

         It is useful to introduce a compact notation for standard
         states: \begin{equation}\label{asbtnot}|\asbt\rangle =
         (\ad_{\alpha})^{s}(\bd_{\beta})^{t}|0\rangle.\end{equation}
         A unitary addition operator, $\tilde{+},$ is defined by
         \begin{equation}\label{defplus}\begin{array}{l}
         \tilde{+}|\alpha s,\beta t\rangle |\alpha^{\p}s^{\p},\beta^{\p}
         t^{\p}\rangle|0\rangle= \\ \hspace{1cm}|\alpha s,\beta t\rangle
         |\alpha^{\p}s^{\p},\beta^{\p} t^{\p}\rangle|\alpha s,
         \beta t+\alpha^{\p} s^{\p},\beta^{\p}t^{\p}\rangle\end{array}\end{equation}
         where \begin{equation}\label{alpbetab}\begin{array}{l}|\alpha s,
         \beta t+\alpha^{\p} s^{\p},\beta^{\p}t^{\p}\rangle \\ \hspace{0.5cm}=(\ad_{\alpha})^{s}
         (\bd_{\beta})^{t}(\ad_{\alpha^{\p}})^{s^{\p}}
         (\bd_{\beta^{\p}})^{t^{\p}}|0\rangle =\\ \hspace{1cm}=(\ad_{\alpha})^{s}
         (\ad_{\alpha^{\p}})^{s^{\p}}(\bd_{\beta})^{t}(\bd_{\beta^{\p}})^{t^{\p}}
         |0\rangle \\ \hspace{1.5cm}=|\alpha s+\alpha^{\p} s^{\p},\beta
         t+\beta^{\p}t^{\p}\rangle .\end{array}\end{equation}
         This result, which uses the commutativity of the $a$ and $b$
         a-c operators, shows the separate addition of the $a$
         and  $b$ components of the states.\footnote{The product state
         representation is used here as it
         is familiar. The three states can be represented in the
         a-c operator formalism as a single state by expanding
         $\mathcal H^{Ra}$ to include operators for three
         distinguishable pairs of distinguishable systems (i.e.
         $(a,b)\rightarrow (a,b),(c,d),(e,f)$ or by adding an
         additional integral index to $a,b$ that has the values
         $1,2,3$ for the three different states in the product.
         In this case there are just two distinguishable systems.}

         This also shows that the result of addition need not be a
         standard representation even if the inputs are standard.
         This is the case if $s$ and $\sp$ or $t$ and $\tp$ have
         common elements or if $\a\neq\ap$ or $\b\neq\bp.$

          The notation of Eq.\ \ref{asbtnot} will be used for both
         fermions and bosons with the understanding that for
         fermions the real component $(\ad_{\a})^{s}(\ad_{\ap})^{\sp}$
         is given by Eq.\ \ref{stdfer}. Also if $\a =\ap$ then for any
         sites $j$ that $s,\sp$ have in common, the operator product
         $\ad_{\a,1,j}\ad_{\a,1,j}$ is replaced by $\ad_{\a,2,j}\ad_{\a,1,j}.$
         Also for fermions the equality sign in Eq.
         \ref{defplus} is replaced by $=_{\pm}$ or equality up to
         the sign. If the number of fermions in $|\asbt
         +\asbtp\rangle$ is odd the sign is minus.  Otherwise it
         is even.\footnote{One way to make the sign always $+$
         is to require that the dynamical steps
         of addition conserve fermion number by use of an
         additional set of fermions to serve as a sink or source.}
         In addition the right hand operator products
         $(\ad_{\a})^{s}(\ad_{\ap})^{\sp}(\bd_{\b})^{t}(\bd_{\bp})^{\tp}$
         must  be written in the ordering given in Eqs. \ref{occferm}
         and \ref{occnoferm}. All the above changes for fermions
         are duplicated for the $\bd$ operator products.

          The operator, $\tilde{+},$  acting on states that are
          linear superpositions of rational states generates
          entanglement. To see this Let $\psi=\sum_{\alpha,
         s,\beta,t}d_{\alpha,s,\beta,t}|\alpha s,\beta t\rangle$
         and $\psi^{\p}=\sum_{\alpha^{\p},
         s^{\p},\beta^{\p},t^{\p}}d^{\p}_{\alpha^{\p},s^{\p},
         \beta^{\p},t^{\p}}|\alpha^{\p} s^{\p},\beta^{\p}
         t^{\p}\rangle.$ Then
          \begin{equation}\label{plusentngl}\begin{array}{l}
          \tilde{+}\psi\,\psi^{\prime}|0\rangle =\sum_{\alpha
          ,s,\beta,t}\sum_{\alpha^{\p},s^{\p},\beta^{\p},t^{\p}}
          d_{\alpha, s,\beta,t}d^{\p}_{\alpha^{\p},
          s^{\p},\beta^{\p},t^{\p}} \\ \hspace{1cm}\times|\alpha s,\beta t\rangle
          |\alpha^{\p}s^{\p},\beta^{\p}t^{\p}\rangle|\alpha s,\beta t +\alpha^{\p}
          s^{\p},\beta^{\p}t^{\p}\rangle\end{array} \end{equation} which is entangled.

          To describe repeated arithmetic operations
          it is useful to have a state that describes the result
          of addition of $\psi$ to $\psi^{\prime}$. This state is
          obtained by taking the trace over the first two
          components of $\tilde{+}\psi\psi^{\p}|0\rangle$ in Eq.
          \ref{plusentngl}:
          \begin{equation}\label{rhoaddcmplx}\begin{array}{l}
          \rho_{\psi+\psi^{\prime}}=Tr_{1,2}\tilde{+}
          |\psi\rangle|\psi^{\p}\rangle|0\rangle\langle 0|\langle\psi^{\p}
          |\langle\psi|\tilde{+}^{\dag}= \sum_{\alpha,\beta, s,t}
          \\ \times\sum_{\alpha^{\p},\beta^{\p},s^{\p},t^{\p}}
          |d_{\alpha, s,\beta,t}|^{2}|d^{\p}_{\alpha^{\p},s^{\p},\beta^{\p},t^{\p}}|^{2}
          \rho_{\alpha s,\beta t+ \alpha^{\p}s^{\p},\beta^{\p}t^{\p}}.\end{array}
          \end{equation} Here $\rho_{\alpha s,\beta t+
          \alpha^{\p}s^{\p},\beta^{\p}t^{\p}}$ is the pure state
          density operator $|\alpha s,\beta
          t+ \alpha^{\p}s^{\p},\beta^{\p}t^{\p}\rangle\langle\alpha s,\beta
          t+ \alpha^{\p}s^{\p},\beta^{\p}t^{\p}|.$  The expectation value
          of $\tilde{N}$ on this state gives
          \begin{equation}\label{Nplus} Tr(\tilde{N}\rho_{\psi+\psi^{\prime}})
          =\langle\psi|\tilde{N} |\psi\rangle+\langle\psi^{\prime}
          |\tilde{N}|\psi^{\prime}\rangle\end{equation} which is
          as expected.

          For subtraction one notes that
          $|\alpha^{\p}s,\beta^{\p}t\rangle$ is the additive
          inverse of $|\alpha s,\beta t\rangle$ if $\alpha^{\p}\neq\alpha$
          and $\beta^{\p}\neq\beta.$ Then \begin{equation}\label{addinv}
          |\alpha s,\beta t+\alpha^{\p}s,\beta^{\p}t\rangle =_{N}|0\rangle
          \end{equation} where Eq. \ref{abdjabj} is used to give $(\ad_{\alpha})^{s}
          (\ad_{\alpha^{\p}})^{s}=_{N}1=_{N}(\bd_{\beta})^{t}
          (\bd_{\beta^{\p}})^{t}.$ This can be used to define a
          unitary subtraction operator $\tilde{-}$ by \begin{equation}\label{subtr}
          \tilde{-}|\asbt\rangle |\asbtp\rangle|0\rangle=
          \tilde{+}|\asbt\rangle|\a^{\p\p}s,\b^{\p\p}t\rangle|0\rangle\end{equation}
          where $\a^{\p\p}\neq\ap$ and $\b^{\p\p}\neq\bp.$

         \section{Summary and Discussion}\label{SD}

         In this paper a quantum mechanical representation of
         complex rational numbers in binary was given that does not
         depend on qubits. Instead binary numbers are represented as a
         distribution of fermion or boson systems along an integer
         lattice that correspond to a distribution of $\pm 1s$ along
         the lattice. The creation operators for bosons are
         $\ad_{\a,j}\bd_{\b,j}.$  For fermions they are $\ad_{\a,h,j}
         \bd_{\b,h,j}.$ Here $\a,\b=+,-$ and $h$ is an additional
         parameter to allow for more than one fermion with the
         same sign and $j$ value.

         Complex rational numbers are represented by states
         given as strings of creation operators acting on the
         vacuum state $|0\rangle.$  Both standard and nonstandard
         representations of numbers occur. Standard representations
         are limited to the states with  operator
         strings that cannot be simplified by use of Eqs.
         \ref{abdjabj} and \ref{abjabj} for bosons, and
         \ref{fabdjabj} and \ref{fabjabj} for fermions and all
         $\ad$ operators have the same sign and all $\bd$
         operators have the same sign. All other basis states
         correspond to nonstandard representations.

         This representation has some advantages over the usual
         qubit representation.  It is compact as it avoids the
         use of $0s$ which serve only as place holders, and is
         especially useful to represent numbers with $1s$
         separated by long strings of $0s$. It is also suitable for
         representation of  columns of positive and negative real
         and imaginary numbers that are to be added together as
         one nonstandard state.

         The representation used here may also suggest new
         physical models for quantum computers.  For bosons an
         example would be a string of four possible types of Bose
         Einstein condensates (BEC) at each lattice site. The four
         types correspond to positive real $(\ad_{+})$, negative
         real $(\ad_{-})$, positive imaginary $(\bd_{+})$,
         negative imaginary $(\bd_{-})$. Values are determined
         by the numbers of each type of system at each site $j$.
         Computation operations correspond to changing the numbers
         of systems at the lattice sites.  A similar
         representation for fermions is possible except that more
         than one fermion of the $a$ type  or of the $b$ type at
         the same lattice location and with
         the same sign would have different $h$ values.

         \section*{Acknowledgements}
          This work was supported by the U.S. Department of Energy,
          Office of Nuclear Physics, under Contract No. W-31-109-ENG-38.

           \end{document}